# Theory of anomalous Hall effect for type-II high-Tc and conventional superconductors

**Ming Ju Chou**[1] **, Wei Yeu Chen**[2,*]

[1] Department of Physics, National Taiwan Normal University, Taipei 10610, Taiwan

[2] Department of Physics, Tamkang University, Tamsui 25137, Taiwan

## ABSTRACT

The anomalous Hall effect for type-II superconductors is investigated by random walk theorem. It is shown that the origin of Hall anomaly is induced by the thermally activated vortex bundle flow (TAVBF) over the directional-dependent energy barrier formed by the Magus force, random collective pinning force, and strong pinning force inside the vortex bundles. The directional-dependent potential barrier of the vortex bundles renormalizes the Hall and longitudinal resistivities strongly. Under the framework of present theory, it is also shown that the Hall anomaly is universal for type-II superconductors, either high-$T_c$ or conventional as well as bulk materials or thin films. The conditions for Hall anomaly and reentry phenomenon are derived, the Hall and longitudinal resistivities as well as Hall angle for type-II superconducting films and bulk materials versus temperature and applied magnetic field are calculated. All the results are in agreement with the experiments.

[*] Corresponding author. Tel.: 886-2-22346169

E-mail: wychen@mail.tku.edu.tw (W.Y. Chen)

# 1 Introduction

The sign reversal of the Hall resistivity below the superconducting transition temperature $T_c$ is the most confusing and controversial phenomenon for the last forty years since it was first observed by Van Beelen et. el. in 1967 [1]. Experimental data show that the Hall resistivity in many high-$T_c$ and conventional type-II superconductors [1-54] exhibits the sign reversal phenomenon, namely, it changes its sign from positive to negative, after reaching a minimum, then increases again as the temperature or applied magnetic field decreases, when the temperature is slightly below the critical temperature of the superconductors [1, 36-38]. In some materials, the observed Hall resistivity exhibits the double sign reversal, or reentry phenomenon [40, 41]. Various theories [4, 27-29], trying to explain this anomaly, have been proposed, such as, the large thermomagnetic theory [27], flux flow theory [28], opposing drift of quasiparticles theory [29], and many others. However, up to the present time, there is no satisfactory explanation. The origin of this Hall anomaly still remains unsolved.

Recently, we [6] have developed a self-consistent theory of thermally activated vortex bundle flow (TAVBF) over the directional-dependent potential barrier by applying the random walk theorem. According to this theory it is shown that the origin of the anomalous Hall effect is induced by the thermally activated vortex bundle flow (TAVBF) over the directional-dependent potential barrier. The directional-dependent energy barrier means that the energy barrier is a function of the direction of thermally activated motion of the vortex bundles. The conditions for appearing of the sign reversal and double sign reversal of the Hall resistivity for type-II superconductors are derived under the framework of present theory. It is shown that the directional-dependent potential barrier of the vortex bundles renormalizes the Hall and longitudinal resistivities strongly. It is also shown that the Hall anomaly is universal for type-II superconductors, either high-$T_c$ or conventional as well as for bulk materials or thin films.

Based on the present theory, the Hall and longitudinal resistivities as well as Hall angle as functions of temperature and applied magnetic field are calculated for type-II superconducting films and bulk materials. All the results are in agreement with the experiments.

The rest of the paper is organized as follows. In section 2, a mathematical description of the model is presented. The theory of anomalous Hall effect is developed in section 3. Within this theory, the anomalous Hall effect for type-II superconducting bulk materials is studied in section 4. In section 5, the corresponding Hall anomaly for superconducting films is investigated. In section 6, the reentry phenomenon and the conditions for occurring of this fascinating phenomenon are analyzed. Finally, several important issues about our theory and concluding remarks are presented in sections 7 and 8, respectively.

## 2 Mathematical Description of the Model

Let us consider a type-II conventional or high-$T_c$ superconductor, the Hamiltonian of the fluctuation for the flux line lattice (FLL) in the $z-direction$ is given by [6-12, 23]

$$H = H_f + H_R \tag{1}$$

where $H_f = H_{kin} + H_e$ represents the Hamiltonian for the free modes [6-12, 23], with $H_{kin}$ the kinetic energy part [6-12, 23]

$$H_{kin} = \frac{1}{2\rho} \sum_{\vec{K}\mu} P_\mu(\vec{K}) P_\mu(-\vec{K}) \tag{2}$$

$H_e$ the elastic energy part [6-13, 23],

$$H_e = \frac{1}{2} \sum_{\vec{K}\mu\nu} C_L K_\mu K_\nu S_\mu(\vec{K}) S_\nu(-\vec{K}) + \frac{1}{2} \sum_{\vec{K}\mu} (C_{66} K_\perp^2 + C_{44} K_z^2) S_\mu(\vec{K}) S_\mu(-\vec{K}) \tag{3}$$

and $H_R$ represents the random Hamiltonian, given as [6-12, 23],

$$H_R = \sum_{\vec{K}\mu} f_{R\mu}(\vec{K}) S_\mu(-\vec{K}) \tag{4}$$

where $(\mu,\nu) = (x,y)$, $\rho$ is the effective mass density of the flux line [14], $K_\perp^2 = K_x^2 + K_y^2$, $P_\mu(\vec{K}), S_\mu(\vec{K})$ are the Fourier transformations of the momentum and displacement operators, and $C_L, C_{11}, C_{44}$ and $C_{66}$ are temperature- and $\vec{K}$-dependent bulk modulus, compression modulus, tilt modulus and shear modulus, respectively [15-19]. $\vec{f}_R(\vec{K})$ is the Fourier transformation of the collective pinning force $\vec{f}_R(\vec{r}) = -\vec{\nabla} V_R(\vec{r})$, with $V_R(\vec{r})$ the random potential energy of the collective pinning [20, 21], which is the sum of the contributions of defects within a distance $\xi$ of the vortex core position $\vec{r}$, where $\xi$ is the temperature-dependent coherent length. The correlation functions of the random collective pinning force are assumed to be the short-range correlation [7],

$$\overline{<< f_{R\alpha}(\vec{k})\, f_{R\beta}^*(\vec{k}') >>_{th}} = \beta^C(T,B)\, \delta_{\alpha\beta}\, \delta(\vec{k}-\vec{k}') \tag{5}$$

where $\overline{<< \quad >>_{th}}$ are the quantum, thermal, and random averages, and $\beta^C(T,B)$ is the temperature- and magnetic field-dependent correlation strength.

The equation of motion of the displacement operator $S_\mu(\vec{K})$ can be obtained from Eq. (1) as

$$\rho \ddot{S}_\mu(\vec{K}) + C_L(\vec{K}\cdot\vec{S}(\vec{K}))K_\mu + (C_{66}K_\perp^2 + C_{44}K_z^2)S_\mu(\vec{K}) + f_\mu(\vec{K}) = 0 \tag{6}$$

Then the solution of Eq. (6) can be obtained as

$$S_\mu(\vec{K}) = S_{R\mu}(\vec{K}) + S_{f\mu}(\vec{K}) \tag{7}$$

where $S_{R\mu}(\vec{K})$ denotes the deformation displacement operator of the FLL due to the collective pinning of the random function $\vec{f}_{R\mu}(\vec{K})$, and $S_{f\mu}(\vec{K})$ the displacement operator for the fluctuation of the free modes. They are given by

$$S_{R\mu}(\vec{K}) = [(\vec{K}\cdot\vec{f}_R(\vec{K}))\frac{\delta_{\alpha,1}}{K_\perp}]\cdot\frac{1}{C_{11}K_\perp^2 + C_{44}K_z^2} + [f_{R\alpha}(\vec{K}) - (\vec{K}\cdot\vec{f}_R(\vec{K}))\frac{\delta_{\alpha,1}}{K_\perp}]\cdot\frac{1}{C_{66}K_\perp^2 + C_{44}K_z^2} \tag{8}$$

and

$$S_{f\mu}(\vec{K}) = \sqrt{\frac{\hbar}{2\rho\omega_{K\mu}}}(\alpha^+_{-\vec{K}\mu} + \alpha_{\vec{K}\mu}) \tag{9}$$

respectively, where $\mu = 1$ presents the component parallel to the $\vec{K}_\perp$ direction, while $\mu = 2$ is perpendicular to the $\vec{K}_\perp$ direction. It is understood that the free Hamiltonian can be diagonalized with the eigenmodes spectrum [6-12, 23]

$$\omega_{K1} = [\frac{1}{\rho}(C_{11}K_\perp^2 + C_{44}K_z^2)]^{\frac{1}{2}}$$

$$\omega_{K2} = [\frac{1}{\rho}(C_{66}K_\perp^2 + C_{44}K_z^2)]^{\frac{1}{2}} \tag{10}$$

with $\alpha^+_{\vec{K}\mu}$, $\alpha_{\vec{K}\mu}$ are the creation and the annihilation operators for the corresponding eigenmodes.

However, the quenched disorder always destroys the long-range order of the FLL, after which only short-range order, the vortex bundle, prevails [5-7]. The corresponding transverse size of vortex bundle $|\vec{R}|$ is determined by the relation [9, 10]

$$\overline{\ll |\vec{S}_R(\vec{R}) - \vec{S}_R(0)|^2 \gg_{th}} = r_f^2 \tag{11}$$

where $r_f$ represents the random collective pinning force range, once again, $\overline{\ll \;\; \gg_{th}}$ are the thermal, quantum and random averages.

## 3 Theoretical investigation of anomalous Hall effect by random walk theorem

The theory of anomalous Hall effect is studied based on the self-consistent theory of thermally activated vortex bundle flow (TAVBF), under the steady-state condition, over the directional-dependent potential barrier via random walk theorem [6] is developed in this section. The corresponding coherent oscillation frequency $\nu_c$ of the vortex bundles and the mean value of the angle between the random collective pinning force of the vortex bundles and positive $y$-direction are evaluated under the conditions $J < J_c$, where $J_C$ is the critical current density of the superconductor. Finally, the temperature- and field-dependent Hall and longitudinal resistivities induced by the vortex bundles flow are calculated.

### 3.1 Coherent oscillation frequency and mean direction of random collective pinning force for vortex bundles

In this subsection let us consider a p-type superconductor, with the applied magnetic field $\vec{B}$ in the z-axis and the external current density $\vec{J}$ in the x-axis $\vec{J} = J\,\vec{e}_x$ with $J < J_C$. The equation of motion of the vortex line inside the vortex bundle driving by the thermal radiation of frequency $\omega$ is given by

$$M_v \frac{d^2\vec{r}_v}{dt^2} = q\,\vec{E}\,e^{i\omega t} + J\,\Phi_0 \vec{e}_x \times \vec{e}_z - (\frac{M_v}{\tau_R}\frac{d\vec{r}_v}{dt}) - (k_R\,\vec{r}_v) + \vec{f}_{el} \tag{12}$$

where $M_v$ is the effective mass of the vortex line, and $\vec{r}_v$ the displacement of vortex line from its equilibrium position. In the first term on the right hand side of Eq. (12), $\vec{E}e^{i\omega t}$ denotes the driving electric field of the thermal radiation for the vortex line, and $q$ stands for the total circulating charge of the vortex line. The second term is the Lorentz force with $\Phi_0$ the unit flux. $1/\tau_R$ characterizes the damping rate associated with the motion of the vortex line, $k_R$ represents the restoring force constant of the vortex line under the action of random collective pinning force, and $\vec{f}_{el}$ is the elastic force of the vortex line inside the vortex bundle. The homogeneous solution of the Eq. (12) vanishes quickly due to the presence of damping. The particular solution includes two parts: the time-dependent and time-independent parts.

The time-dependent part of the particular solution oscillates with frequency $\omega$ about a new equilibrium position, which is determined by the time-independent part of the particular solution. By identifying the oscillation energy of the vortex line inside the potential barrier with the thermal energy, the thermal oscillation frequency $\nu$ of the individual vortex inside the potential barrier can be expressed as

$$\nu = \bar{\nu}\sqrt{T} \tag{13}$$

with $\bar{\nu} = (1/\pi A)\sqrt{k_B/2M_v}$ , where $A$ stands for the random- and thermal-averaged amplitude of the oscillation of the vortex line in the bundle, and $k_B$ the Boltzmann constant.

However, the oscillations of vortex lines inside the vortex bundle are not coherent, namely, their oscillations are at random. To obtain the coherent oscillation frequency $\nu_c$ of the vortex bundle as a whole, by utilizing the random walk's theorem, the frequency $\nu$ in Eq. (13) must be divided by the square root of N, the number of vortices inside the vortex bundle

$$\nu_c = \frac{\nu}{\sqrt{N}} = \frac{\bar{\nu}\sqrt{T}\sqrt{\Phi_0}}{R\sqrt{\pi B}} \tag{14}$$

where $R$ is the transverse size of the vortex bundle and $B$ the value of the applied magnetic field.

The time-independent part of the particular solution is give by

$$\vec{r}_p = \frac{-J\,\Phi_0\,\vec{e}_y + \vec{f}_{el}}{k_R} \tag{15}$$

The above result indicates that the vortex line moves to a new equilibrium position $\vec{r}_p$ from its original one. Since the elastic force $\vec{f}_{el}$ is much less than the Lorentz force $J\Phi_0$ , the angle $\theta$ between the random collective pinning force and the positive y-direction measured in the counterclockwise sense can be obtained approximately as

$$\theta \cong \frac{|\vec{f}_{el}|}{|\vec{f}_L|} = \frac{|\vec{f}_{el}|}{J\,\Phi_0} \tag{16}$$

where $|\vec{f}_{el}|$ and $|\vec{f}_L|$ are the magnitudes of elastic force and Lorentz force of the vortex line, respectively. Taking into account the fact that the compression modulus $C_{11}$ is much larger than shear modulus $C_{66}$ [7, 8], we arrive at, the magnitude of the displacement vector $|\vec{S}_f(\vec{r})|$ of the vortex line as well as its corresponding elastic force $|\vec{f}_{el}|$ is proportional to $\sqrt{k_B/C_{66}}$ , or $(1/\sqrt{B})\sqrt{T/(T_C-T)}$ .

The temperature- and field-dependent $\theta(T,B)$ can therefore be written as

$$\theta(T,B) = \alpha\frac{1}{\sqrt{B}}\sqrt{\frac{T}{T_c-T}} \tag{17}$$

where $\alpha$ is a proportional constant. By applying the random walk theorem, the mean angle $\Theta(T,B)$ between the random collective pinning force of vortex bundle and positive y-direction measures in counterclockwise sense, can be expressed as

$$\Theta(T,B) = \sqrt{N}\,\theta(T,B) = \bar{\alpha}\,\sqrt{\frac{T}{T_c - T}} \tag{18}$$

where $\bar{\alpha} = \alpha\, R\sqrt{\pi/\Phi_0}$ .

3.2 Directional-dependent energy barrier

In this subsection we shall calculate the directional-dependent energy barrier of the vortex bundles generated by the Magnus force, random collective pinning force, and the strong pinning force inside the vortex bundle. The directional-dependent potential barrier means that the energy barrier is a function of direction of the thermally activated motion. Assuming that the external current $J < J_c$, and is in the x-direction and the applied magnetic field $\vec{B}$ in the z-direction, the Magnus force $\vec{F}_M$ acting on the vortex bundle can then be obtained as

$$\vec{F}_M = \bar{V}\, n_s\, e\, (\vec{v}_T - \vec{v}_b) \times B\,\hat{e}_z \tag{19}$$

where $\vec{v}_b$ is the velocity of the thermally activated vortex bundle flow, $\bar{V}$ is the volume of the vortex bundle, $e$ is the electron charge, $n_s$ and $\vec{v}_T$ are the supercharge density and its velocity, respectively, combining with $\vec{J} = n_s\, e\vec{v}_T = J\,\vec{e}_x$, and $\vec{B} = B\vec{e}_z$. From the theory of mechanics, the potential generated by a force field $\vec{F}(\vec{r})$ can be expressed as

$$V(\vec{r}) - V(0) = -\int_0^{\vec{r}} \vec{F}(\vec{r}) \cdot d\,\vec{r} \tag{20}$$

After some algebra, the directional-dependent energy barrier of the vortex bundles both in the positive and negative x-direction as well as y-direction are obtained as

$$U + \bar{V}R\,(JB\frac{v_{by}}{v_T} - \langle F_{p_x} \rangle_R)$$

$$U - \bar{V}R\,(JB\frac{v_{by}}{v_T} - \langle F_{p_x} \rangle_R)$$

$$U + \bar{V}R\,(JB - JB\frac{v_{bx}}{v_T} - \langle F_{p_y} \rangle_R)$$

$$U - \bar{V}R\,(JB - JB\frac{v_{bx}}{v_T} - \langle F_{p_y} \rangle_R) \tag{21}$$

respectively, where $U$ is the potential barrier generated by the strong pinning force due to the randomly distributed strong pinning sites inside the vortex bundle, $R$ represents the transverse size of the vortex bundle, the range of $U$ is assumed to be of the order $R$, and $<\vec{F}_p>_R$ stands for the random average of the random collective pinning force per unit volume.

### 3.3 Vortex bundle flow velocity and its induced longitudinal and Hall resistivities

The results of Eq. (21) indicate that the directional-dependent potential barrier used for calculating the vortex bundle flow velocity actually itself contains the vortex bundle flow velocity. Therefore, the velocity of thermally activated vortex bundle flow (TAVBF) over the directional-dependent energy barrier must be solved self-consistently as follows

$$v_{by} = \nu_c R\{\exp[\frac{-1}{k_B T}(U + \bar{V} R(J B - J B\frac{v_{bx}}{v_T} - < F_{p_y} >_R))]$$
$$- \exp[\frac{-1}{k_B T}(U - \bar{V} R(J B - J B\frac{v_{bx}}{v_T} - < F_{p_y} >_R))]\} \quad (22)$$

$$v_{bx} = \nu_c R\{\exp[\frac{-1}{k_B T}(U + \bar{V} R(J B\frac{v_{by}}{v_T} - < F_{p_x} >_R))]$$
$$- \exp[\frac{-1}{k_B T}(U - \bar{V} R(J B\frac{v_{by}}{v_T} - < F_{p_x} >_R))]\} \quad (23)$$

with $\nu_c$ is the coherent oscillation frequency of the vortex bundle. Taking into account the fact that $(v_{bx}/v_T) << 1$, from the above equations, the vortex bundle flow velocity in the y and x components can be approximately obtained as

$$v_{by} = (\frac{\bar{V}}{R}\sqrt{\frac{T\,\Phi_0}{\pi B}})\, R\, \exp(\frac{-U}{k_B T})\{\exp[\frac{-\bar{V} R}{k_B T}(J B - | < F_p >_R | \cos\Theta)]$$
$$- \exp[\frac{+\bar{V} R}{k_B T}(J B - | < F_p >_R | \cos\Theta)]\} \quad (24)$$

$$v_{bx} = (\frac{\bar{V}}{R}\sqrt{\frac{T\,\Phi_0}{\pi B}})\, R\, \exp(\frac{-U}{k_B T})\, \{\exp[\frac{-\bar{V} R}{k_B T}(\frac{-J B | v_{by} |}{v_T} + | < F_p >_R | \sin\Theta)]$$
$$- \exp[\frac{+\bar{V} R}{k_B T}(\frac{-J B | v_{by} |}{v_T} + | < F_p >_R | \sin\Theta)]\} \quad (25)$$

respectively, where $\Theta(T,B)$ is the mean angle between the random collective pinning force of the vortex bundles and positive $y$-direction measured in the counterclockwise sense. By considering the identities $\vec{E} = -\vec{v}_b \times \vec{B}$, $\rho_{xx} = E_x / J$, $\rho_{xy} = E_y / J$, together with Eq. (18) and bearing in mind that $\Theta(T,B)$ is usually very small, then the longitudinal and Hall resistivities induced by the vortex bundle flow can now be obtained, respectively, as follows:

$$\rho_{xx} = \frac{\bar{v}\sqrt{BT\Phi_0}}{J\sqrt{\pi}} \exp(\frac{-U}{k_B T})\{\exp[\frac{\bar{V}R}{k_B T}(JB - (\frac{\beta^C(T,B)}{\bar{V}})^{\frac{1}{2}})]$$
$$-\exp[\frac{-\bar{V}R}{k_B T}(JB - (\frac{\beta^C(T,B)}{\bar{V}})^{\frac{1}{2}})]\} \tag{26}$$

$$\rho_{xy} = \frac{-\bar{v}\sqrt{BT\Phi_0}}{J\sqrt{\pi}} \exp(\frac{-U}{k_B T})\{\exp[\frac{\bar{V}R}{k_B T}((\frac{\beta^C(T,B)}{\bar{V}})^{\frac{1}{2}}\bar{\alpha}\sqrt{\frac{T}{T_C - T}} - JB\frac{|v_{by}|}{v_T})]$$
$$-\exp[\frac{-\bar{V}R}{k_B T}((\frac{\beta^C(T,B)}{\bar{V}})^{\frac{1}{2}}\bar{\alpha}\sqrt{\frac{T}{T_C - T}} - JB\frac{|v_{by}|}{v_T})]\} \tag{27}$$

and

$$|v_{by}| = J\rho_{xx} / B \tag{28}$$

with $(\beta^C(T,B)/\bar{V})^{1/2}$ is the magnitude of the random average of the random collective pinning force per unit volume, and $\bar{\alpha}$ is a proportional constant. The results of Eqs. (26) and (27) demonstrate that the directional-dependent potential barrier of the vortex bundle renormalizes the Hall and longitudinal resistivities strongly. It is well understood that when temperature (applied magnetic field) below $T_1(B_1)$, the quasiorder-disorder first-order phase transition temperature (magnetic field) [7, 8, 11, 12], the vortex lines form large vortex bundles of dimension $R \approx 10^{-6} m$; while temperature (applied magnetic field) above $T_1(B_1)$, the vortex lines form a disordered amorphous vortex system , however, they are not randomly distributed single-quantized vortex lines. In other words, the vortex lines still bounded together to form small vortex bundles of dimension $R \approx 10^{-8} m$ [11, 12]. Therefore, taking into account the fact that the arguments in the exponential functions inside the curly bracket of Eqs. (26) and (27) are very small when the Lorentz force is close to the random collective pinning force, we finally obtain the temperature- and field- dependent longitudinal resistivity $\rho_{xx}$, Hall resistivity $\rho_{xy}$ and Hall angle $\theta_H$ as

$$\rho_{xx} = \frac{\bar{v}\sqrt{B\Phi_0}}{J\sqrt{\pi T}} \exp(\frac{-U}{k_B T})(\frac{2\bar{V}R}{k_B})[JB - (\frac{\beta^C(T,B)}{\bar{V}})^{\frac{1}{2}})] \tag{29}$$

$$\rho_{xy} = \frac{-\bar{v}\sqrt{B\Phi_0}}{J\sqrt{\pi T}} \exp(\frac{-U}{k_B T})(\frac{2\bar{V}R}{k_B})[(\frac{\beta^C(T,B)}{\bar{V}})^{\frac{1}{2}}\bar{\alpha}\sqrt{\frac{T}{T_C - T}} - JB\frac{|v_{by}|}{v_T}] \tag{30}$$

$$\theta_H = \tan^{-1}(\frac{\rho_{xy}}{\rho_{xx}}) = \tan^{-1}[\frac{JB(|v_{by}|/v_T) - (\beta^C(T,B)/\bar{V})^{\frac{1}{2}}\bar{\alpha}\sqrt{T/(T_C - T)}}{JB - (\beta^C(T,B)/\bar{V})^{\frac{1}{2}}}] \tag{31}$$

respectively, with $|v_{by}| = J\rho_{xx}/B$. The above results give rise to the anomalous Hall effect. The conditions for the appearance of anomalous Hall effect can be derived as follows: Let us assume that the vortex system is initially in the region of thermally activated motion of small vortex bundles. From our previous study [6], the value of $(1/B)(\beta^C(T,B)/\bar{V})^{1/2}$ increases with decreasing applied magnetic field when the temperature $T$ of the system is kept at a constant value; if this value is greater than the value of $(|v_{by}|/v_T)/(\bar{\alpha}\sqrt{T/(T_C - T)})$, then the Hall resistivity $\rho_{xy}$ in Eq. (30) changes its sign from positive to negative with decreasing applied magnetic field. On the other hand, the value of $(1/B)(\beta^C(T,B)/\bar{V})^{1/2}$ increases and that of $\bar{\alpha}\sqrt{T/(T_C - T)}$ decreases as temperature decreasing when the applied magnetic field $B$ is kept at a constant value. Therefore, the term $(1/B)(\beta^C(T,B)/\bar{V})^{1/2}\bar{\alpha}\sqrt{T/(T_C - T)}$ exists a maximum at some temperature $T$ if this maximum is greater than $J\ (|v_{by}|/v_T)$, then $\rho_{xy}$ possesses the sign reversal property. According to the above analysis, the condition for appearing the Hall anomaly is

$$\frac{1}{B}\left(\frac{\beta^C(T,B)}{\bar{V}}\right)^{\frac{1}{2}}\bar{\alpha}\sqrt{\frac{T}{T_c - T}} > J\frac{|v_{by}|}{v_T} \tag{32}$$

The present results demonstrate that the anomalous Hall effect is induced by the thermally activated vortex bundle flow (TAVBF) over the directional-dependent energy barrier. Moreover, $\rho_{xy}$ might possibly possess the double sign reversal property or reentry phenomenon at temperature $T = T_R$, the reentry temperature. In other words, $\rho_{xy}$ initially decreases, crossing over from positive to negative, after reaching a minimum, then increases again crossing over back from negative to positive at $T = T_R$. This fascinating phenomenon of reentry will be presented in detail in Section 6.

Next let us investigate the behavior of longitudinal resistivity $\rho_{xx}$ as functions of temperature and applied magnetic field. As we have mentioned above, the value of $(1/B)(\beta^C(T,B)/\bar{V})^{1/2}$ increases with decreasing temperature (applied magnetic field); therefore, $\rho_{xx}$ in Eq. (29) decreases monotonically as temperature (applied magnetic field) decreasing. The first-order phase transition between the small and large vortex bundles of the vortex system takes place [7, 8] when temperature (applied magnetic field) below $T_1(B_1)$, the vortex system then crosses over to the large vortex bundles. In this case, the potential barrier $U$ generated by the strong pinning force due to the randomly distributed strong pinning sites inside the vortex bundle will become large; in other words, both of the Hall and longitudinal resistivities will decay to zero quickly with decreasing temperature (applied magnetic field).

It is worthwhile to point out that all the above derivations are valid for type-II conventional and high-$T_c$ superconductors. Although their mechanisms [30-35] and the method of pairing are different, these only affect the structure of the vortex lattice and is a minute effect in our theory. It is also interesting to note that dimensional fluctuations only have an effect on the volume of short-range order, and do not affect the essential structure of our theory. Therefore, the above formalisms are valid both for superconducting bulk materials and thin films. Within the framework of present theory, it is demonstrated that the Hall anomaly is universal for type-II conventional and high-$T_c$ as well as for bulk materials and thin films superconductors provided certain conditions are satisfied. Detailed calculations and discussions will be presented in subsequent sections.

## 4 Anomalous Hall effect for type-II superconducting bulk materials

In this section first let us concentrate on the case for type-II superconducting bulk materials. The volume $\bar{V}$ for the vortex bundle in Eqs. (29) and (30) is given as $\bar{V} = \pi R^2 L$, where $R$ and $L$ are the transverse and longitudinal sizes of the vortex bundle, respectively. In this case, the longitudinal and Hall resistivities for type-II superconducting bulk materials now become

$$\rho_{xx} = \frac{\bar{v}\sqrt{\Phi_0}\sqrt{B}}{J\sqrt{\pi}\sqrt{T}} \exp(\frac{-U}{k_B T}) \, [\frac{2\pi R^3 L}{k_B}][J B - (\frac{\beta^C(T,B)}{\bar{V}})^{\frac{1}{2}}] \quad (33)$$

$$\rho_{xy} = \frac{-\bar{v}\sqrt{\Phi_0}\sqrt{B}}{J\sqrt{\pi}\sqrt{T}} \exp(\frac{-U}{k_B T}) \, [\frac{2\pi R^3 L}{k_B}] \, [(\frac{\beta^C(T,B)}{\bar{V}})^{\frac{1}{2}} \bar{\alpha} \sqrt{\frac{T}{T_c - T}} - J B \frac{|v_{by}|}{v_T}] \quad (34)$$

respectively, with $|v_{by}| = J\rho_{xx}/B$. The Eq. (34) indicates that the value of Hall resistivity $\rho_{xy}$ changes its sign from positive to negative with decreasing applied magnetic field (temperature) if the condition for appearing Hall anomaly given by Eq. (32) $(1/B)(\beta^C(T,B)/\bar{V})^{1/2}\bar{\alpha}\sqrt{T/(T_c - T)} > J(|v_{by}|/v_T)$ is satisfied. The situations for constant temperature and constant applied magnetic field of this Hall anomaly are investigated separately as follows.

### 4.1 Hall anomaly for constant temperature

Under the framework of present theory, the Hall resistivity $\rho_{xy}$ in Eq. (34) does change its sign from positive to negative as the applied magnetic field decreasing when the temperature of the system is kept at a constant value $T$. The numerical calculation results of $\rho_{xy}$ and $\rho_{xx}$ of Eqs. (33) and (34) when the temperature is kept at $T = 91K$, with B the applied magnetic field in Tesla are given as follows:

$\rho_{xy}(B=3.5) = 1.2914\times10^{-9}\,\Omega - m$, $\rho_{xy}(3.03) = 1.7658\times10^{-15}\,\Omega - m$, $\rho_{xy}(2.5) = -1.4627\times10^{-9}\,\Omega - m$,

$\rho_{xy}(2) = -2.7907\times10^{-9}\,\Omega - m$, $\rho_{xy}(1.5) = -3.9996\times10^{-9}\,\Omega - m$, $\rho_{xy}(1) = -4.6405\times10^{-9}\,\Omega - m$,

$\rho_{xy}(0.75) = -3.9801\times10^{-9}\Omega - m$, $\rho_{xy}(0.5) = -2.0098\times10^{-9}\Omega - m$; $\rho_{xx}(3.5) = 1.8739\times10^{-6}\Omega - m$, $\rho_{xx}(3.03) = 1.5916\times10^{-6}\Omega - m$, $\rho_{xx}(2.5) = 1.2663\times10^{-6}\Omega - m$, $\rho_{xx}(2) = 9.5142\times10^{-7}\Omega - m$, $\rho_{xx}(1.5) = 6.2539\times10^{-7}\Omega - m$, $\rho_{xx}(1) = 2.9669\times10^{-7}\Omega - m$, $\rho_{xx}(0.75) = 1.6825\times10^{-7}\Omega - m$, $\rho_{xx}(0.5) = 1.226\times10^{-7}\Omega - m$.

The above results show that $\rho_{xy}$ decreases initially, crossing over from positive to negative near 3.03 Tesla, after reaching a minimum at 1 Tesla, then increases again, while $\rho_{xx}$ decreases monotonically as the applied magnetic field decreasing. As we have mentioned in the last section, the first-order phase transition between the small and large vortex bundles of the vortex system occurs [7, 8] when the applied magnetic field decreases below the first-order phase transition magnetic field $B_1 = 0.5$ Tesla. In this case, the vortex system crosses over from small vortex bundles to large vortex bundles, the potential barrier $U$ produced by the randomly distributed strong pinning sites inside the vortex bundle becomes large; hence, both of the Hall and longitudinal resistivities induced by the vortex bundles flow approach to zero quickly with decreasing applied magnetic field. These results are in agreement with the experimental data on $YBa_2Cu_3O_{7-\delta}$ high-$T_c$ bulk materials [36]. In obtaining the above results, the following approximate data have been employed:

$R = 2\times10^{-8}m$, $L = 10^{-6}m$, $J = 10^6 A/m^2$, $T_c = 92K$, $v_T = 10^3 m/\sec$, $\bar{\nu} = 10^{11}\sec^{-1}$, $\bar{\alpha} = 5.59\times10^{-5}T^{\frac{-1}{2}}$, $\exp(\frac{-U}{k_B T}) = 2.07\times10^{-2}m$, $(\frac{\beta^C(B=3.5)}{\bar{V}})^{\frac{1}{2}} = 3.4506\times10^6 N/m^3$,

$(\frac{\beta^C(3.03)}{\bar{V}})^{\frac{1}{2}} = 2.9849\times10^6 N/m^3$, $(\frac{\beta^C(2.5)}{\bar{V}})^{\frac{1}{2}} = 2.4605\times10^6 N/m^3$,

$(\frac{\beta^C(2)}{\bar{V}})^{\frac{1}{2}} = 1.9668\times10^6 N/m^3$, $(\frac{\beta^C(1.5)}{\bar{V}})^{\frac{1}{2}} = 1.4748\times10^6 N/m^3$,

$(\frac{\beta^C(1)}{\bar{V}})^{\frac{1}{2}} = 9.854\times10^5 N/m^3$, $(\frac{\beta^C(0.75)}{\bar{V}})^{\frac{1}{2}} = 7.4042\times10^5 N/m^3$,

and $(\frac{\beta^C(0.5)}{\bar{V}})^{\frac{1}{2}} = 4.915\times10^5 N/m^3$.

### 4.2 Hall anomaly for constant magnetic field

Under the framework of present theory, the results of numerical calculations of the Hall and longitudinal resistivities induced by the thermally activated vortex bundle flow (TAVBF) of small vortex bundles when the applied magnetic field is kept at a constant value $B = 2.24$ Tesla are given as follows:

$\rho_{xy}(T = 91.6K) = 6.96\times10^{-10}\Omega - m$, $\rho_{xy}(91.3) = -1.82\times10^{-11}\Omega - m$, $\rho_{xy}(91) = -2.491\times10^{-9}\Omega - m$, $\rho_{xy}(90) = -4.199\times10^{-9}\Omega - m$, $\rho_{xy}(89) = -4.722\times10^{-9}\Omega - m$, $\rho_{xy}(88) = -3.081\times10^{-9}\Omega - m$;

$\rho_{xx}(91.6)=1.878\times10^{-6}\Omega-m$, $\rho_{xx}(91.3)=1.396\times10^{-6}\Omega-m$, $\rho_{xx}(91)=1.094\times10^{-6}\ \Omega-m$,

$\rho_{xx}(90)=7.028\times10^{-7}\Omega-m$, $\rho_{xx}(89)=5.22\times10^{-7}\Omega-m$, $\rho_{xx}(88)=4.91\times10^{-7}\Omega-m$.

The above results indicate that $\rho_{xy}$ decreases initially, crossing over from positive to negative near 91.3 K, after reaching a minimum at about 89 K, then increases again, while $\rho_{xx}$ decreases monotonically as decreasing the temperature. As we have noted in last section, the first-order phase transition between the small and large vortex bundles of the vortex system takes place [7, 8] when the temperature decreases below the first-order phase transition temperature $T_1=88K$ between the small and large vortex bundles. In this case the vortex system crosses over from small vortex bundles to large vortex bundles, and the potential barrier $U$ becomes large, then, both of Hall and longitudinal resistivities induced by the vortex bundles flow reduce to zero quickly with decreasing temperature. These results are in agreement with the experimental data on $YBa_2Cu_3O_{7-\delta}$ high-$T_c$ bulk materials [36]. In arriving at the above results, the following approximate data have been used:

$R=2\times10^{-8}m$, $L=10^{-6}m$, $J=10^6A/m^2$, $T_c=92K$, $v_T=10^3m/\sec$, $\bar{v}=10^{11}\sec^{-1}$,

$\bar{\alpha}=5.59\times10^{-5}T^{\frac{-1}{2}}$, $\exp(\frac{-U}{k_BT})=2.07\times10^{-2}$, $(\frac{\beta^C(T=91.6)}{\bar{V}})^{\frac{1}{2}}=2.178\times10^6N/m^3$,

$(\frac{\beta^C(91.3)}{\bar{V}})^{\frac{1}{2}}=2.194\times10^6N/m^3$, $(\frac{\beta^C(91)}{\bar{V}})^{\frac{1}{2}}=2.204\times10^6N/m^3$,

$(\frac{\beta^C(90)}{\bar{V}})^{\frac{1}{2}}=2.217\times10^6N/m^3$, $(\frac{\beta^C(89)}{\bar{V}})^{\frac{1}{2}}=2.223\times10^6N/m^3$,

and $(\frac{\beta^C(88)}{\bar{V}})^{\frac{1}{2}}=2.224\times10^6N/m^3$.

## 5 Anomalous Hall effect for type-II superconducting films

Now let us turn our attention to type-II superconducting thin films, the corresponding volume $\bar{V}$ of the vortex bundle is therefore given by $\bar{V}=\pi R^2d$, where $R$ is the transverse size of the vortex bundle and $d$ the thickness of the film. The longitudinal and Hall resistivities of Eqs. (29) and (30) can now be described, respectively, as

$$\rho_{xx}=\frac{\bar{v}\sqrt{\Phi_0}\sqrt{B}}{J\sqrt{\pi}\sqrt{T}}\exp(\frac{-U}{k_BT})[\frac{2\pi R^3d}{k_B}][JB-(\frac{\beta^C(T,B)}{\bar{V}})^{\frac{1}{2}}] \tag{35}$$

$$\rho_{xy}=\frac{-\bar{v}\sqrt{\Phi_0}\sqrt{B}}{J\sqrt{\pi}\sqrt{T}}\exp(\frac{-U}{k_BT})\,[\frac{2\pi R^3d}{k_B}]\,[(\frac{\beta^C(T,B)}{\bar{V}})^{\frac{1}{2}}\,\bar{\alpha}\sqrt{\frac{T}{T_c-T}}-J\,B\frac{|v_{by}|}{v_T}] \tag{36}$$

with $|v_{by}|=J\rho_{xx}/B$. The above equations give rise to the phenomenon of anomalous Hall effect for both constant applied magnetic field as well as constant temperature, if the condition for appearing Hall anomaly given by Eq. (32) $(1/B)(\beta^C(T,B)/\bar{V})^{1/2}\bar{\alpha}\sqrt{T/(T_c-T)} > J(|v_{by}|/v_T)$ is satisfied. All the results will be discussed separately as follows.

5.1 Hall anomaly for constant temperature

Based on the present theory, the numerical calculation results of Hall resistivity $\rho_{xy}$ and longitudinal resistivity $\rho_{xx}$ induced by the thermally activated vortex bundle flow (TAVBF) as functions of applied magnetic field in Tesla for the small vortex bundles when temperature is kept at a constant value $T=4.5K$ are given as follows:

$\rho_{xy}(B=7.5)=4.3399\times10^{-11}\Omega-m$, $\rho_{xy}(7.25)=1.5804\times10^{-11}\Omega-m$, $\rho_{xy}(7)=-2.6307\times10^{-11}\ \Omega-m$,

$\rho_{xy}(6.75)=-6.7189\times10^{-11}\Omega-m$, $\rho_{xy}(6.5)=-1.023\times10^{-10}\Omega-m$, $\rho_{xy}(6.25)=-1.283\times10^{-10}\Omega-m$,

$\rho_{xy}(6)=-1.1842\times10^{-10}\Omega-m$, $\rho_{xy}(5.75)=-8.8118\times10^{-11}\Omega-m$, $\rho_{xy}(5.5)=-3.752\times10^{-11}\Omega-m$;

$\rho_{xx}(7.5)=8.2782\times10^{-7}\Omega-m$, $\rho_{xx}(7.25)=7.8079\times10^{-7}\Omega-m$, $\rho_{xx}(7)=7.2721\times10^{-7}\Omega-m$,

$\rho_{xx}(6.75)=6.742\times10^{-7}\Omega-m$, $\rho_{xx}(6.5)=6.24\times10^{-7}\Omega-m$, $\rho_{xx}(6.25)=5.78\times10^{-7}\Omega-m$,

$\rho_{xx}(6)=5.492\times10^{-7}\Omega-m$, $\rho_{xx}(5.75)=5.3044\times10^{-7}\Omega-m$, $\rho_{xx}(5.5)=5.2209\times10^{-7}\Omega-m$.

The above results show that $\rho_{xy}$ decreases initially, crossing over from positive to negative near 7.15 Tesla, after reaching a minimum at near 6.25 Tesla, then increases again, while $\rho_{xx}$ decreases monotonically as decreasing the applied magnetic field. As we have mentioned before, the first-order phase transition between the small and large vortex bundles of the vortex system takes place [7, 8] when the applied magnetic field decreases below the first-order phase transition magnetic filed $B_1=5.5$ Tesla between the small and large vortex bundles. In this case, the potential barrier $U$ generated by the strong pinning force due to the randomly distributed strong pinning sites inside the vortex bundle becomes large, therefore, both of $\rho_{xy}$ and $\rho_{xx}$ reduce to zero rapidly. The above results are in agreement with experimental data on $Mo_3Si$ conventional low-$T_c$ superconducting films [37]. In obtaining the above results, the following approximate data have been employed:

$R=2\times10^{-8}m$, $d=5\times10^{-8}m$, $J=1.5\times10^5 A/m^2$, $T_c=7.5K$, $v_T=30\,m/\sec$,

$\bar{\alpha}=1.0449\times10^{-3}T^{\frac{-1}{2}}$, $\bar{\nu}=10^{11}\sec^{-1}$, $\exp(\frac{-U}{k_B T})=3.0899\times10^{-4}$,

$(\frac{\beta^C(B=7.5)}{\bar{V}})^{\frac{1}{2}}=4.5772\times10^5 N/m^3$, $(\frac{\beta^C(7.25)}{\bar{V}})^{\frac{1}{2}}=4.4737\times10^5 N/m^3$,

$(\frac{\beta^C(7)}{\bar{V}})^{\frac{1}{2}}=4.4324\times10^5 N/m^3$, $(\frac{\beta^C(6.75)}{\bar{V}})^{\frac{1}{2}}=4.3964\times10^5 N/m^3$,

$(\frac{\beta^C(6.5)}{\bar{V}})^{\frac{1}{2}} = 4.3483\times10^5 N/m^3$, $(\frac{\beta^C(6.25)}{\bar{V}})^{\frac{1}{2}} = 4.272\times10^5 N/m^3$,

$(\frac{\beta^C(6)}{\bar{V}})^{\frac{1}{2}} = 4.0516\times10^5 N/m^3$, $(\frac{\beta^C(5.75)}{\bar{V}})^{\frac{1}{2}} = 3.7418\times10^5 N/m^3$,

and $(\frac{\beta^C(5.5)}{\bar{V}})^{\frac{1}{2}} = 3.335\times10^5 N/m^3$.

5.2 Hall anomaly for constant magnetic field

The numerical calculations of $\rho_{xy}$ and $\rho_{xx}$ induced by the thermally activated vortex bundle flow (TAVBF) for small vortex bundles when the applied magnetic field is kept at a constant value $B = 2$ Tesla are given as follows:

$\rho_{xy}(T = 92K) = 2.0023\times10^{-9}\,\Omega - m$, $\rho_{xy}(91.5) = 4.9915\times10^{-10}\,\Omega - m$, $\rho_{xy}(91) = -1.252\times10^{-9}\,\Omega - m$, $\rho_{xy}(90.5) = -2.4984\times10^{-9}\,\Omega - m$, $\rho_{xy}(90) = -2.902\times10^{-9}\,\Omega - m$, $\rho_{xy}(89.5) = -2.0005\times10^{-9}\,\Omega - m$, $\rho_{xy}(89) = -7.5142\times10^{-10}\,\Omega - m$; $\rho_{xx}(92) = 9.7317.\times10^{-7}\,\Omega - m$, $\rho_{xx}(91.5) = 8.6241\times10^{-7}\,\Omega - m$, $\rho_{xx}(91) = 7.7545\times10^{-7}\,\Omega - m$, $\rho_{xx}(90.5) = 7.082\times10^{-7}\,\Omega - m$ , $\rho_{xx}(90) = 6.5791\times10^{-7}\,\Omega - m$, $\rho_{xx}(89.5) = 6.2461\times10^{-7}\,\Omega - m$, $\rho_{xx}(89) = 5.9962\times10^{-7}\,\Omega - m$.

The above results show that $\rho_{xy}$ decreases initially, crossing over from positive to negative near 91.5 K, after reaching a minimum at about 90 K, then increases again, while $\rho_{xx}$ decreases monotonically with decreasing the temperature. As we have mentioned before, the first-order phase transition between the small and large vortex bundles of the vortex system occurs [7, 8] when the temperature decreases below the first-order phase transition temperature $T_1 = 89\,K$ between the small and large vortex bundles. In this case, the potential barrier $U$ becomes large; namely, both of $\rho_{xy}$ and $\rho_{xx}$ decrease quickly to zero with deceasing temperature. These results are in agreement with the experimental data on $YBa_2Cu_3O_7$ high-$T_c$ superconducting thin films [38]. In obtaining the above results, the following approximate data have been used:

$R = 2\times10^{-8}m$, $d = 5\times10^{-8}m$, $J = 10^6 A/m^2$, $T_c = 94K$, $v_T = 10^2 m/\sec$,

$\bar{\alpha} = 7.325\times10^{-4}T^{\frac{-1}{2}}$, $\bar{\nu} = 10^{11}\sec^{-1}$, $\exp(\frac{-U}{k_B T}) = 2.01\times10^{-1}$,

$(\frac{\beta^C(T = 92)}{\bar{V}})^{\frac{1}{2}} = 1.9298\times10^6 N/m^3$, $(\frac{\beta^C(91.5)}{\bar{V}})^{\frac{1}{2}} = 1.938\times10^6 N/m^3$,

$(\frac{\beta^C(91)}{\bar{V}})^{\frac{1}{2}} = 1.9444\times10^6 N/m^3$, $(\frac{\beta^C(90.5)}{\bar{V}})^{\frac{1}{2}} = 1.9493\times10^6 N/m^3$,

$(\frac{\beta^C(90)}{\bar{V}})^{\frac{1}{2}} = 1.9531\times10^6 N/m^3$, $(\frac{\beta^C(89.5)}{\bar{V}})^{\frac{1}{2}} = 1.9556\times10^6 N/m^3$,

and $(\frac{\beta^C(89)}{\overline{V}})^{\frac{1}{2}} = 1.9575\times10^6 N/m^3$ .

## 6 Reentry phenomenon

Experimental data show that for some materials, the observed Hall resistivity exhibits the double sign reversal or reentry phenomenon. Based on the discussions in Section 3, the conditions for the occurrence of this fascinating reentry phenomenon can be obtained as

$$\frac{1}{B}(\frac{\beta^C(T,B)}{\overline{V}})^{\frac{1}{2}}\overline{\alpha}\sqrt{\frac{T}{T_c-T}} > J\frac{|v_{by}|}{v_T} \qquad \text{and} \qquad T_1 < T_R \tag{37}$$

where $T_1$ is the temperature of the first-order phase transition between the small and large vortex bundles of the vortex system, and $T_R$ is the reentry temperature, namely, the temperature for the value of $\rho_{xy}$ crosses over back from negative to positive. Taking into account the fact that $T_R$ increases with increasing $(\beta^C(T,B)/\overline{V})^{1/2}$, the random collective pinning force per unit volume of the vortex system [39]. Therefore, the reentry phenomenon could be observed for materials with larger random collective pinning force, such as, $YBa_2Cu_3O_{7-\delta}$ [40] or $Tl_2Ba_2Cu_2O_8$ [41].

The numerical calculations of Hall resistivity $\rho_{xy}$ and longitudinal resistivity $\rho_{xx}$ induced by the thermally activated vortex bundle flow (TAVBF) as functions of temperature for small vortex bundles when the applied magnetic field is kept at a constant value $B = 2$ Tesla are given as follows:

$\rho_{xy}(T=102K) = 2.1346\times10^{-11}\Omega-m$ , $\rho_{xy}(100) = -3.69\times10^{-19}$ $\Omega-m$ , $\rho_{xy}(98) = -1.4067\times10^{-11}\Omega-m$ ,

$\rho_{xy}(96) = -2.3518\times10^{-11}\Omega-m$ , $\rho_{xy}(92) = -1.03436\times10^{-11}\Omega-m$ , $\rho_{xy}(88) = 5.9409\times10^{-19}\Omega-m$ ,

$\rho_{xy}(84) = 5.46\times10^{-12}$ $\Omega-m$ , $\rho_{xy}(78) = 1.3011\times10^{-11}\Omega-m$ , $\rho_{xy}(76) = 1.2967\times10^{-11}\Omega-m$ ;

$\rho_{xx}(102) = 1.6728\times10^{-8}\Omega-m$ , $\rho_{xx}(100) = 1.0657\times10^{-8}$ $\Omega-m$ , $\rho_{xx}(98) = 7.4819\times10^{-9}\Omega-m$ ,

$\rho_{xx}(96) = 5.4754\times10^{-9}\Omega-m$ , $\rho_{xx}(92) = 5.16087\times10^{-9}\Omega-m$ , $\rho_{xx}(88) = 5.1382\times10^{-9}\Omega-m$ ,

$\rho_{xx}(84) = 4.951\times10^{-9}\Omega-m$ , $\rho_{xx}(78) = 4.8488\times10^{-9}\Omega-m$ , $\rho_{xx}(76) = 4.65\times10^{-9}\Omega-m$ .

It is shown that, for constant applied magnetic field $B = 2$ Tesla, $\rho_{xy}$ decreases initially, crossing over from positive to negative near 100 K, after reaching a minimum at 96 K, then increases crossing over back from negative to positive near 88K , reaching a local maximum at about 78K, then decreases again, while $\rho_{xx}$ decreases monotonically as decreasing the temperature. As we have noted before, the first-order phase transition between the small and large vortex bundles of the vortex system takes place [7, 8] when the temperature decreases below the first-order phase transition temperature $T_1 = 76\ K$ between the small and large vortex bundles. The vortex system then crosses over from the small to large vortex bundles. In this case, the potential barrier $U$ produced by the strong pinning force due to the randomly distributed strong pinning sites inside the vortex bundle becomes large; hence, both of $\rho_{xy}$ and $\rho_{xx}$ decrease promptly to zero with decreasing temperature. The above results are in agreement with the experimental data on $Tl_2Ba_2Cu_2O_8$ high-$T_c$ superconducting films [41]. In obtaining the above results, the following approximate data have been used:

$R = 2\times 10^{-8} m$, $d = 10^{-6} m$, $J = 10^7 A/m^2$, $T_c = 104K$, $v_T = 10^2 m/\sec$,

$\bar{\alpha} = 1.12\times 10^{-4} T^{\frac{-1}{2}}$, $\bar{v} = 10^{11} \sec^{-1}$, $\exp(\frac{-U}{k_B T}) = 8.3199\times 10^{-5}$,

$(\frac{\beta^C (T=102)}{\bar{V}})^{\frac{1}{2}} = 1.8464\times 10^7 N/m^3$, $(\frac{\beta^C (100)}{\bar{V}})^{\frac{1}{2}} = 1.9031\times 10^7 N/m^3$,

$(\frac{\beta^C (98)}{\bar{V}})^{\frac{1}{2}} = 1.93267\times 10^7 N/m^3$, $(\frac{\beta^C (96)}{\bar{V}})^{\frac{1}{2}} = 1.9512\times 10^7 N/m^3$,

$(\frac{\beta^C (92)}{\bar{V}})^{\frac{1}{2}} = 1.955\times 10^7 N/m^3$, $(\frac{\beta^C (88)}{\bar{V}})^{\frac{1}{2}} = 1.95618\times 10^7 N/m^3$,

$(\frac{\beta^C (84)}{\bar{V}})^{\frac{1}{2}} = 1.9588\times 10^7 N/m^3$, $(\frac{\beta^C (78)}{\bar{V}})^{\frac{1}{2}} = 1.961069\times 10^7 N/m^3$,

and $(\frac{\beta^C (76)}{\bar{V}})^{\frac{1}{2}} = 1.9631\times 10^7 N/m^3$.

## 7 Discussions

In this section we would like to point out first of all that under the present theory the anomalous Hall effect is induced by the thermally activated vortex bundle flow (TAVBF) over the directional-dependent potential barrier formed by the Magnus force and the random collective pinning force as well as the potential barrier generated by the strong pinning force due to the randomly distributed strong pinning sites inside the vortex bundle if certain given conditions are satisfied.

Secondly, it is worthwhile to note that our theory for the anomalous Hall effect is a microscopic theory. After the theoretical woks of Ginzburg, Landau [52], Abrikosov [2] and Gor'kov [53], the flux line in the mixed state type-II superconductors form a long-range order of the vortex line lattice. However, the existing theories in the literature on vortex dynamics are classical where the kinetic energy term in the Hamiltonian is omitted. Recently, we [6-12, 23] have developed a quantum theory for the vortex dynamics with full Hamiltonian including the kinetic energy, elastic energy and random energy. The presence of random energy due to quenched disorder, the long-range order of the FLL is destroyed, only the short-range order prevails [6-12, 23]. By applying the random walk theorem, we arrive at the theory of thermally activated vortex bundle flow (TAVBF) over the directional-dependent potential barrier occurs for certain ranges of temperature and applied magnetic filed for $J < J_c$. Therefore, the present theory for the anomalous Hall effect is indeed a microscopic theory.

Thirdly, we only consider the systems that are in steady state, any time-dependent behavior of the system will not be discussed.

Fourthly, our theory is a very general one. It can be applied to type-II conventional and high-$T_c$ superconductors. Although their mechanisms, the structure of vortex lattice, and even the method of pairing are entirely different, these do not affect the results of our theory.

Finally, the dimensional fluctuations do not affect the essential structure of our theory. Therefore we could investigate 2D and 3D anomalous Hall effects at the same time.

## 8 Conclusion

In this paper, a microscopic theory of anomalous Hall effect in type-II conventional and high-$T_c$ superconductors via random walk theorem is developed. Based on the present theory, it is shown that the Hall anomaly is induced by the thermally activated vortex bundle flow (TAVBF) over the directional-dependent potential barrier formed by Magnus force, random collective pinning force, and strong pinning force inside the vortex bundles. The conditions for the sign reversal and double sign reversal of the Hall resistivity are obtained. It is shown that the Hall anomaly is universal for type-II superconductors, either conventional or high-$T_c$ as well as for bulk materials and thin films, provided certain given conditions are satisfied. It is also demonstrated that the directional-dependent potential barrier of the vortex bundles renormalizes the Hall and longitudinal resistivities strongly. The Hall and longitudinal resistivities are calculated for type-II superconducting bulk materials as well as thin films as functions of temperature and applied magnetic field. It is shown that they indeed exhibit the anomalous properties, namely, $\rho_{xy}$ decreases initially, crossing over from positive to negative, after reaching a minimum, then increases and approaches to zero, while $\rho_{xx}$ decreases monotonically with decreasing the temperature (applied magnetic field). Finally, the reentry phenomenon is discussed, namely, for some materials when the applied magnetic field is kept at a constant value, the Hall resistivity $\rho_{xy}$ decreases initially, crossing over from positive to negative, after reaching a minimum, then increases crossing over back from negative to positive, after reaching a local maximum, then decreases again as decreasing the temperature. All the results are in agreement with the experiments.

## Acknowledgements

The authors would like to thank Professor E H Brandt for useful and constructive discussions.